\documentclass[aps, prb, twocolumn, superscriptaddress, showpacs, floatfix]{revtex4-2}

\usepackage{amsmath,amsbsy,amssymb,amsfonts}
\usepackage{microtype}

\usepackage{graphicx, color}
\usepackage{natbib}
\usepackage{hyperref}
\usepackage[capitalise]{cleveref}
\usepackage{subfigure}
\usepackage{float}

\usepackage{soul,xcolor}

\newcommand{\eps}{\epsilon}

\newcommand{\tb}[1]{{}{#1}}

\begin{document}
\setstcolor{red}
\graphicspath{{Figures/}}
\date{\today}

\title{Nonlinear Wavepacket Dynamics in Proximity to a Stationary Inflection Point}

\author{Serena Landers}
\affiliation{Wave Transport in Complex Systems Lab, Department of Physics, Wesleyan University, Middletown, CT-06459, USA}
\author{Arkady Kurnosov}
\email{akurnosov@wesleyan.edu}
\affiliation{Wave Transport in Complex Systems Lab, Department of Physics, Wesleyan University, Middletown, CT-06459, USA}
\author{William Tuxbury}
\affiliation{Wave Transport in Complex Systems Lab, Department of Physics, Wesleyan University, Middletown, CT-06459, USA}
\author{Ilya Vitebskiy}
\affiliation{Air Force Research Laboratory, Wright-Patterson AFB, OH, USA}
\author{Tsampikos Kottos}
\affiliation{Wave Transport in Complex Systems Lab, Department of Physics, Wesleyan University, Middletown, CT-06459, USA}

\begin{abstract}
A stationary inflection point (SIP) in the Bloch dispersion relation of a periodic waveguide is an exceptional point degeneracy where three Bloch eigenmodes coalesce forming the so-called frozen mode with a divergent amplitude and vanishing group velocity of its propagating component. We have developed a theoretical framework to study the time evolution of wavepackets centered at an SIP. Analysis of the evolution of statistical moments distribution of linear pulses shows a strong deviation from the conventional ballistic wavepacket dynamics in dispersive media. The presence of nonlinear interactions dramatically changes the situation, resulting in a mostly ballistic propagation of nonlinear wavepackets with the speed and even the direction of propagation essentially dependent on the wavepacket amplitude. Such a behavior is unique to nonlinear wavepackets centered at an SIP {and can be used for the realization of a novel family of beam power routers for classical waves}.
\end{abstract}
\maketitle

\section{Introduction}
The Bloch dispersion relation of a periodic waveguide can develop exceptional points of degeneracy (EPD), where two or more Bloch eigenmodes coalesce. {As opposed to well-studied resonant EPDs, which require the implementation of dissipative mechanisms, Bloch EPDs occur even in the absense of gain/loss elements since they occur in the spectrum of transfer matrices. These are non-Hermitian operators (they are pseudo-unitary, belonging to the $SU(N)$ group), allowing the formation of EPDs in their spectrum. A well-known example is a regular band edge where two counter-propagating Bloch modes collapse onto each other.} 

Our investigation focuses on a stationary inflection point (SIP), where three Bloch eigenmodes (two evanescent and one propagating) coalesce (see \cite{Figotin2006,Figotin2011,Li2017,Tuxbury2022,Nada2021,Furman2023,HerreroParareda2022,Figotin2003} and references therein). In proximity to the SIP frequency, an incident wave can be completely converted into the frozen mode with diverging amplitude and vanishing group velocity of its propagating component \cite{Figotin2011,Li2017,Tuxbury2022,Figotin2003,Ballato2005}. The frozen mode regime is quite different from a common cavity resonance because its frequency is independent of the system dimensions and boundary conditions. The most remarkable features of the frozen mode regime include robustness with respect to structural imperfections and moderate losses \cite{Li2017,Tuxbury2022,Tuxbury2021,Gan2019}. The above properties make the frozen mode regime particularly attractive for the enhancement of various {wave}-matter interactions and {wave} amplification, {including} cavity-less lasing \cite{Ramezani2014,Yazdi2017,HerreroParareda2023}. 

The focus of this study is the unique dynamics of an SIP-centered wavepacket inside a periodic structure. Unlike the monochromatic frozen mode which involves non-Bloch Floquet eigenmodes \cite{Figotin2006,Figotin2011,Li2017,Tuxbury2022,Nada2021,Furman2023,HerreroParareda2022,Figotin2003}, the Gaussian wavepacket is a superposition of propagating Bloch modes with wavenumbers close to that of the SIP. Due to the SIP proximity, both the group velocity and its first derivative with respect to the Bloch wavenumber are infinitesimally small. As a consequence, both linear and nonlinear dynamics of an SIP-centered wavepacket demonstrate some interesting and unique features. Indeed, in the linear regime, the time evolution of the SIP-centered wavepacket does not involve ballistic propagation, which can be expected due to the zero group velocity at the SIP frequency. Remarkably, though, the presence of nonlinearity changes the situation dramatically. We show that the SIP-centered nonlinear wavepackets can propagate ballistically with the speed and even direction of propagation essentially dependent on the wavepacket amplitude. \tb{This feature can potentially be used for the realization of a novel class of beam power routers whose implementation spans a variety of wave frameworks, ranging from photonic metamaterials \cite{Sukhorukov2008, Gutman2012,Gutman2011,Ramezani2014} to phononics and elastodynamic composite media \cite{Chen2021, Wang2022,Bossart2023}.}

{The remainder of the paper is organized as follows. The next section is devoted to establishing a linear model in the context of coupled mode theory and developing a general theoretical framework for describing SIP wavepacket dynamics. \cref{Sec:OnsiteNonLin} discusses the impact of nonlinearities on the crossover of wavepacket time evolution from SIP dynamics to ballistic propagation. Finally, in \cref{Sec:Protocols} we introduce protocols for controlling propagation direction based on input signal amplitude.}

\begin{figure}[htbp]
\center\includegraphics[scale = .4]{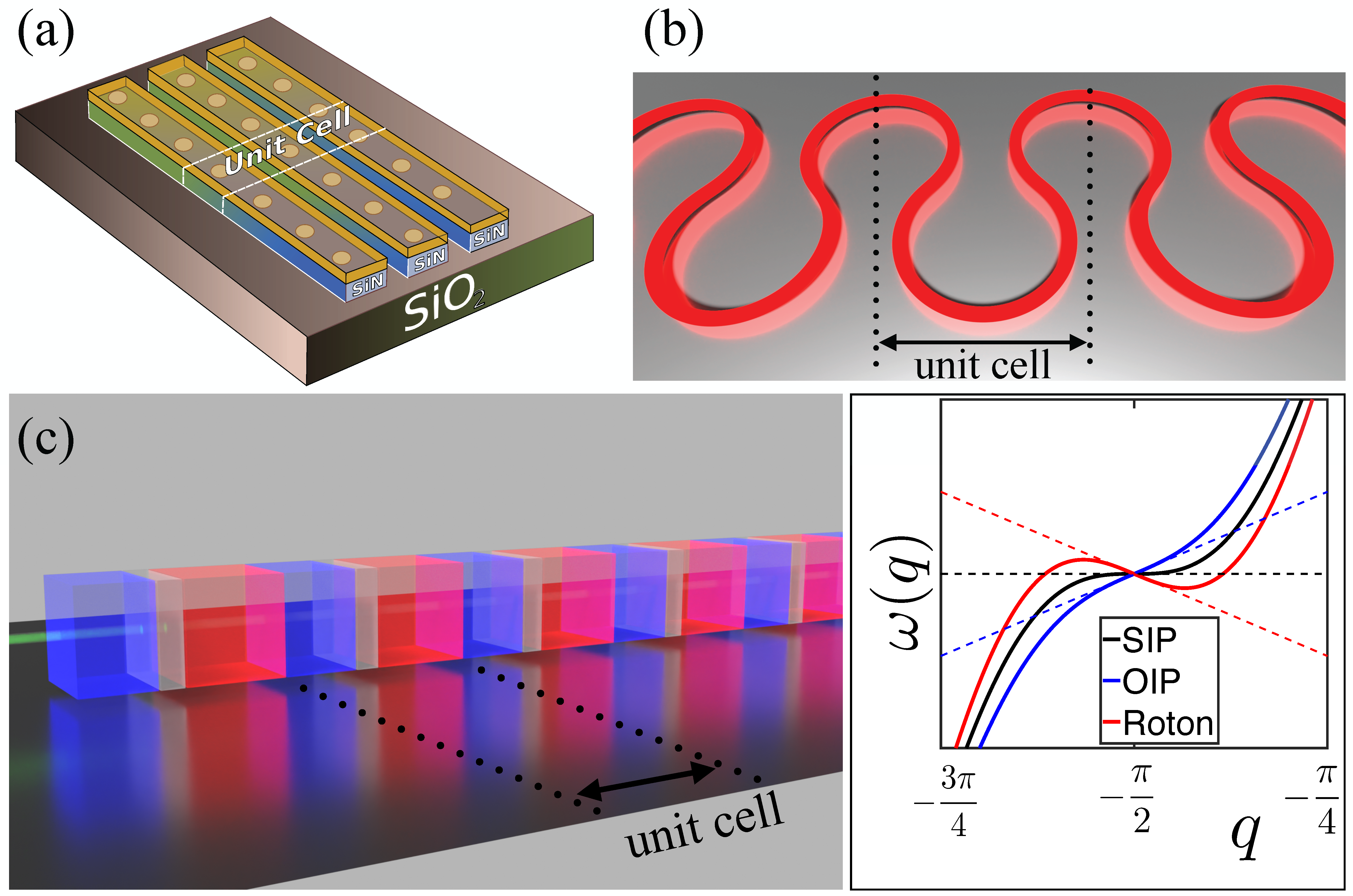}
\caption{\label{Fig:IntroFigure}(a)-(c) Physical systems exhibiting SIP. {The unit cells in each case are indicated with dotted lines}: 
(a) Multimode waveguide array \cite{Sukhorukov2008, Gutman2012,Gutman2011}. (b) Serpentine waveguide \cite{HerreroParareda2022}. (c) Multilayered photonic 
structure \cite{Ramezani2014}. (d) {Dispersion relations, $\omega(q)$, as defined by \cref{Eq:SIPspectrum}, in vicinity 
of inflection point $\bar{q} = -\pi/2$ and the corresponding slopes: black solid for SIP ($\omega^{\prime}(\bar{q}) = \omega^{\prime\prime}(\bar{q}) 
= 0$, $J_{3} = J/3$), blue solid for the ordinary inflection point (OIP), ($\omega^{\prime}(\bar{q})> 0$, $\omega^{\prime\prime}(\bar{q})=0$, $J_{3} 
< J/3$), and red solid line for roton dispersion relation ($\omega^{\prime}(\bar{q})<0$, $\omega^{\prime\prime}(\bar{q})=0$, $J_{3} > J/3$).}}
\end{figure}
\section{Linear Dynamics} For demonstration purposes we consider a minimal mathematical model which may support an SIP. It is provided by the temporal coupled mode theory (CMT) equations
\begin{equation}\label{Eq:DynamicsX-spaceLinear}
i\frac{d\psi_{n}}{dt}  = - J\left(\psi_{n+1} + \psi_{n-1}\right) - J_{3}\left(\psi_{n+3} + \psi_{n-3}\right),\end{equation} where $\psi_n(t)$ is the field amplitude at mode (site) $n = 1, \dots, N$. This model captures all of the features of SIP dynamics, and it can be also associated with a phenomenological description of a physical system. 

{Some of the known examples of photonic setups which exhibit SIPs are illustrated in \cref{Fig:IntroFigure}. The example of multimode waveguide arrays shown in \cref{Fig:IntroFigure}~(a) consists of 3 periodic nanobeams with the same longitudinal periodicity; the possible longitudinal shifts between the waveguides allow for adjustments to the dispersion \cite{Sukhorukov2008, Gutman2012,Gutman2011}. An asymmetric optical periodic serpentine waveguide is presented in \cref{Fig:IntroFigure}~(b); the degree to which the glide symmetry is slightly broken determines the dispersion and can create an SIP \cite{HerreroParareda2022}. Another example of a photonic setup is multilayered photonic structures \cref{Fig:IntroFigure}~(c). The unit cell consists of three components: a central magnetic layer sandwiched between two misaligned anisotropic birefringent layers (blue and red), and the dispersion is controlled by the misalignment angle \cite{Ramezani2014}. SIP-based systems can be also implemented in the acoustic metamaterial framework \cite{Chen2021, Wang2022,Bossart2023}. There is strong consensus in the scientific literature that the primary qualitative features of the SIP-related frozen mode regime remain the same regardless of the specific physical platform. }

 In the case that \cref{Eq:DynamicsX-spaceLinear} describes a set of $N$ coupled resonators with (third-)nearest neighbor coupling constant $(J_3) J$, the variable $t$ indicates time. The same equation might also be used to describe the paraxial field propagation in multicore optical fibers. In this case, $t$ describes the paraxial propagation distance.

Dynamical equations \cref{Eq:DynamicsX-spaceLinear} can be generated by classical Hamiltonian 
\begin{multline}\label{Eq:Hamiltonian1D}
H = \sum\limits_{n}\Bigl[\varepsilon|\psi^{}_{n}|^{2} + J\left(\psi^{}_{n}\psi_{n+1}^{\ast} + c. c.\right) +\\ J_{3} \left(\psi_{n}^{}\psi_{n+3}^{\ast}+ c. c.\right)\Bigr],
\end{multline}
where $\psi^{}_{n}$, $i\psi^{\ast}_{n}$ are canonically conjugate dynamical variables.
Assuming periodic boundary conditions, $\psi_{n+N} = \psi_{n}$,  one can rewrite the Hamiltonian as \begin{equation}\label{Eq:Hamiltonian1Dk-spaceAppendix}
H = \sum_{k}\omega(q_{k})|\phi_{k}|^{2}. \end{equation}
The Bloch modes, $\phi^{}_{k}$ (and their canonically conjugate $i\phi^{\ast}_{k}$), are defined by the Fourier transform
\begin{equation}\label{Eq:FTAppendix}
\psi_{n} = \frac{1}{\sqrt{N}}\sum\limits_{k=-N/2}^{N/2-1}e^{-iq_{k}n}\phi_{k}, \, \phi_{k} = \frac{1}{\sqrt{N}}\sum\limits_{n}e^{iq_{k}n}\psi_{n},
\end{equation}
where $q_{k} = 2\pi k/N$, with a spectrum 
\begin{equation}\label{Eq:SIPspectrum}
\omega(q_{k}) = \varepsilon + 2J\cos q_{k} + 2J_{3}\cos3q_{k}.
\end{equation} 
One can see that for $J_{3}  = J/3$ the dispersion relation exhibits SIPs at $\bar{q} = \pm\pi/2$, as $\omega^{\prime}(\pm\pi/2) = \omega^{\prime\prime}(\pm\pi/2) = 0$, while $\omega^{(3)}(\pm\pi/2) = \mp 8J$. {Of course, the dispersion relation Eq. (\ref{Eq:SIPspectrum}) is associated with a specific mathematical model. In this respect the model Eq. (\ref{Eq:DynamicsX-spaceLinear}) with this dispersion relation is not interesting on its own but rather, it serves as a typical example for presentation purposes and has been used to numerically confirm our general theory for beam dynamics. The theory utilizes only the generic form that the dispersion relation has when expanded around the SIP (see Eq. (\ref{Eq:OmegaExpansionSIP}) below). In this respect, the wavepacket dynamics generated in the proximity of an SIP is indeed universal and model-independent.}

In Bloch mode representation \cref{Eq:DynamicsX-spaceLinear} becomes decoupled,
\begin{equation}\label{Eq:DynamicsK-spaceLinear}
i\frac{d\phi_{k}}{dt} = -\omega(q_{k})\phi_{k}.
\end{equation} In the present study we always assume for the initial condition preparation of a Gaussian packet in $q$-space, \begin{equation}\label{Eq:InitialCondK-space}\phi_{k}(0) = \bar{\phi}_{k} = \left(\frac{4\pi}{N^{2}\sigma^{2}}\right)^{1/4}\exp\left\{-\frac{(q_{k} - \bar{q})^{2}}{2\sigma^{2}}\right\},\end{equation} where we assume the packet to be well confined to the first Brillouin zone, $\sigma\ll 2\pi$, and $\bar{q}$ is the reciprocal lattice vector associated with one of the SIPs. Such initial condition implies a preparation of a Gaussian packet of width $\sigma^{-1}$ in direct space. Assuming the initial wavepacket is centered at $n = 0$,  the time-dependence of the amplitude on the $n$-th site is given by \begin{equation}\label{Eq:psinLinear} \psi_{n}(t) = \frac{\sqrt{N}}{2\pi}\int\limits_{-\pi}^{+\pi}dq e^{iqn}e^{-i\omega(q)t}\bar{\phi}(q), \end{equation} where we have exploited the solution $\phi_{k}(t) = \exp\left\{i\omega(q_{k}) t\right\}\bar{\phi}_{k}$ of \cref{Eq:DynamicsK-spaceLinear} and the condition $N\gg 1$ for continuous limit. 

The integral in \cref{Eq:psinLinear} can be evaluated analytically by employing a number of reasonable approximations. First, the fast convergence of the integral may be exploited by replacing the limits of integration: $\pm\pi\to\pm\infty$. Second, we can use a Taylor expansion of the dispersion relation $\omega(q)$, \cref{Eq:SIPspectrum}, in vicinity of $\bar{q}$, for which the cubic nature of the SIP gives \begin{equation}\label{Eq:OmegaExpansionSIP} \omega(q) \approx \omega_{0} + \frac{\alpha}{3}(q - \bar{q})^{3}, \, \alpha = \frac{1}{2}\frac{d^{3}\omega}{dq^{3}}\Big|_{q = \bar{q}}.\end{equation} A virtue of this approximation transcends a mathematical simplification. Indeed, after utilizing it,  the validity of theoretical conclusions are independent of peculiarities present in the specific model  \cref{Eq:DynamicsX-spaceLinear},  as \cref{Eq:OmegaExpansionSIP} is applicable for any system featuring SIPs. Moreover, the dynamics are determined only by the parameters $\alpha$ and $\sigma$. For the present model,  the parameters we have introduced are given by  $\omega_{0} = \omega(\bar{q}) = \varepsilon \equiv 0$, $\alpha = 8J$, for $\bar{q} = -\pi/2$. 

\begin{figure}[htbp]
\center\includegraphics[scale = .3]{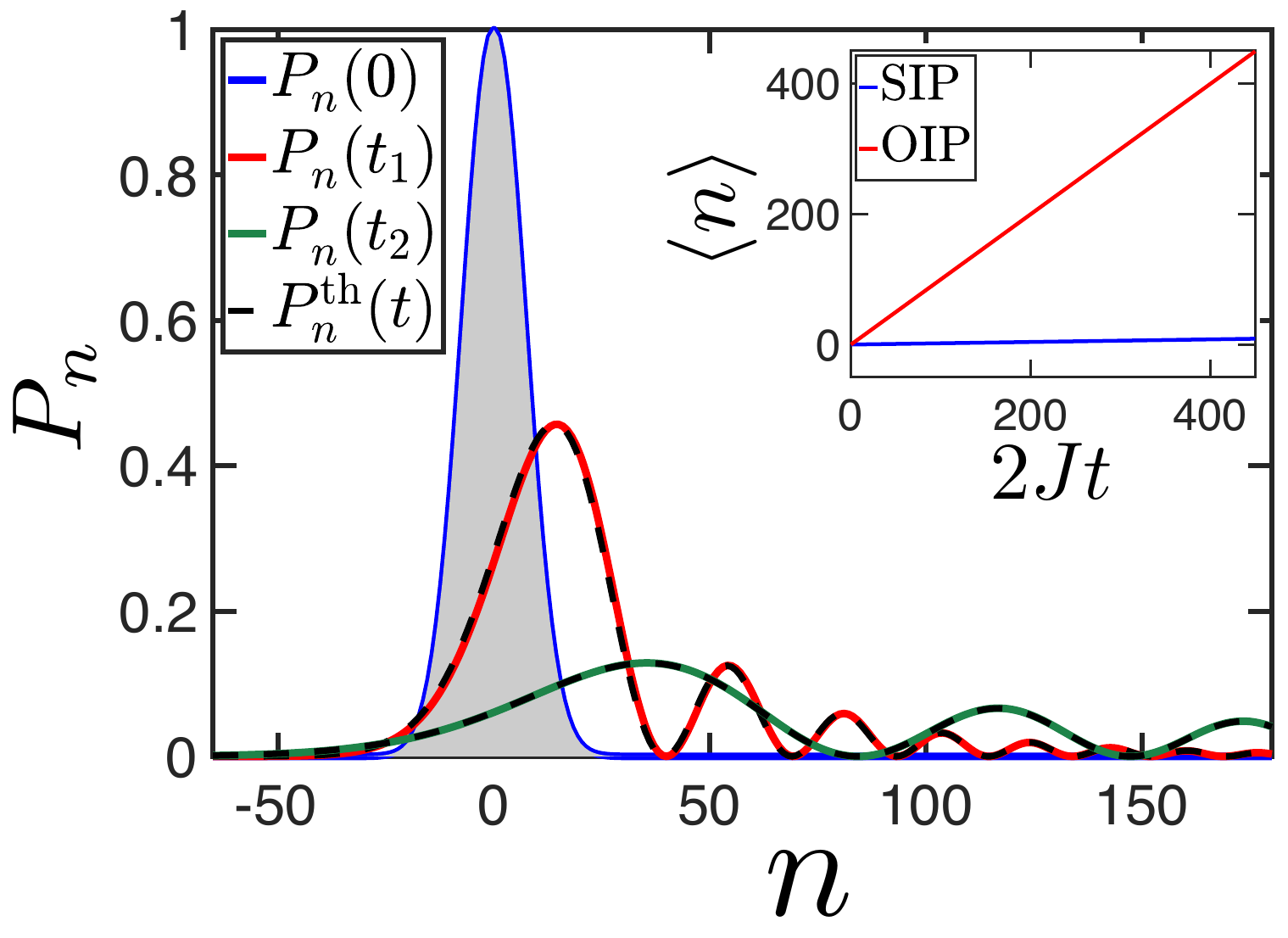}
\caption{\label{Fig:PvsX-MeanVSt}Signal intensity in direct space in presence of an SIP. $P_{n}(0)$ (blue solid line) is the initial wavepacket, $P_{n}(t_{1, 2})$ is the numerical solution of \cref{Eq:DynamicsX-spaceLinear} for $t_{1} < t_{2}$ (red and green solid lines, respectively), $P_{n}^{\rm th}(t)$ (dashed black lines) is given by \cref{Eq:SIPsolution}. Inset: first moment of distribution, $\langle n\rangle$ vs time in presence of an SIP (blue solid line) and an OIP (red solid line); the slopes represent the flow, $\mathcal{F}_{\rm OIP}/\mathcal{F}_{\rm SIP} = 50$. The wavepacket width in $q$-space is $\sigma = 0.1$.}
\end{figure}

Using the approximations we have introduced one can rewrite \cref{Eq:psinLinear} as 
\begin{multline}\label{Eq:psinstep2}
\psi_{n}(t) = \sqrt{2}\pi^{-3/4}\sigma^{-1/2}(\alpha t)^{-1/3}e^{-i\omega_{0}t}e^{i\bar{q}n}\times \\
\int\limits_{0}^{\infty}dx e^{-\epsilon x^{2}}\cos\left(\frac{x^{3}}{3} - zx\right),
\end{multline}
where $z = n(\alpha t)^{-1/3}$, $\epsilon = (1/2)(\alpha t)^{-2/3}\sigma^{-2}$. It can be shown (see \cref{Sec:IntegralAppendix}) that for $t\gg \alpha^{-1}(2\sigma^{2})^{3/2}$, the intensity $P_{n}(t) = |\psi_{n}(t)|^{2}$ on the $n$-th site takes the approximate form, \begin{equation}\label{Eq:SIPsolution}
P_{n}(t) = 2\sqrt{\frac{\pi}{\sigma^{2}}}(\alpha t)^{-2/3}e^{-\frac{n}{\alpha\sigma^{2}t}}\mathrm{Ai}^{2}\left[-n(\alpha t)^{-1/3}\right],
\end{equation}
where $\mathrm{Ai}(-z)$ is Airy function \cite{AbramowitzStegun}. The numerically-evaluated intensity using \cref{Eq:DynamicsX-spaceLinear} as a function of position is reported in \cref{Fig:PvsX-MeanVSt} for $t = 0$ and $t_{2}>t_{1}\gg 0$ (solid lines), while the black dashed lines correspond to the analytical expression \cref{Eq:SIPsolution}.   

The solution we have derived corresponds to a forward propagation of the wavepacket as $P_{n}(t)$ decays quickly for $n<0$ due to the asymptotic behavior of the Airy function; it would be the opposite direction had we prepared the initial packet at the symmetric position in $q$-space, i.~e. at $+\pi/2$, where $\alpha<0$.

To characterize the wavepacket propagation, a good observable is the energy flow, $\mathcal{F}(t) = \sum_{n}n\dot{P}_{n}$, which is equivalent to a time-derivative of the first moment, $\langle n(t)\rangle$. Using \cref{Eq:SIPsolution} one can find the flow of the linear SIP dynamics to be   $\mathcal{F}_{\rm SIP} = {\sigma^{2}\alpha}/{2}$ (see \cref{Sec:FlowAppendix} for mathematical details). 

The same result can be obtained by observing an equality of the flow to the average group velocity, 
\begin{equation}\label{Eq:FlowGroupVel}
\mathcal{F}(t) = \langle v_{g}(t)\rangle = \int dq\left(\frac{\partial\omega}{\partial q}\right)|\phi(q, t)|^{2}.
\end{equation}
This equation remains a good approximation in the presence of weak nonlinearity (see \cref{Sec:FlowVgProofAppendix} for the derivation), and will be helpful for an explanation of a transition to ballistic transport and other nonlinear dynamical effects.

It is possible to consistently single out the anomalous transport features associated with the presence of the SIP in the framework  of the present model. Assuming the long range coupling in \cref{Eq:DynamicsX-spaceLinear}  to be zero, $J_{3} = 0$, then $q=-\pi/2$ corresponds to an ordinary inflection point (OIP), $\omega^{\prime\prime}(q = -\pi/2)=0$, $\omega^{\prime}(q = -\pi/2)\neq 0$, such that the linear term of the $\omega(q)$-expansion dominates in its vicinity, as opposed to the cubic as was the case for an SIP. Therefore, \cref{Eq:psinLinear} is evaluated using the expansion $\omega(q) \approx \omega_{0} + v(q +\pi/2)$ instead of \cref{Eq:OmegaExpansionSIP}, where $v = \omega^{\prime}(q = -\pi/2) = 2J$ is the group velocity. Under the conditions of an OIP, integration of \cref{Eq:psinLinear} results in a direct space Gaussian packet of width $\sigma^{-1}$, propagating at constant velocity $v$. It can be shown that the flow associated with an OIP,  $\mathcal{F}_{\rm OIP} = \langle v_{g}\rangle =v$, does not depend on the initial wavepacket width, $\sigma$, which constitutes  ballistic propagation as opposed to SIP transport. Using explicit expressions for $v$ and $\alpha$ we can see $\mathcal{F}_{\rm SIP}/\mathcal{F}_{\rm OIP} = 2\sigma^{2}$. For instance, the value of $\sigma = 0.1$ used in our numerical simulations implies a 50-fold reduction in the propagation speed of due to a deformation of the dispersion relation towards an SIP (see inset in \cref{Fig:PvsX-MeanVSt}). 
 
\section{Nonlinear dynamics and ballistic crossover \label{Sec:OnsiteNonLin}} As we have established a theoretical framework for slow wave dynamics 
in linear systems that exhibit an SIP, we may explore how this framework is influenced by the presence of weak nonlinearity. {Probably, 
the most common type of nonlinearity is a uniform Kerr-type contribution to the onsite optical potential, which we introduce in the model  by adding the 
term $-\chi|\psi_{n}|^{2}\psi_{n}$ into rhs \cref{Eq:DynamicsX-spaceLinear}, where the nonlinear coefficient $\chi$ could be either positive or negative (focusing/defocusing Kerr nonlinearity)}.

\begin{figure}[htbp]
\center\includegraphics[scale = .25]{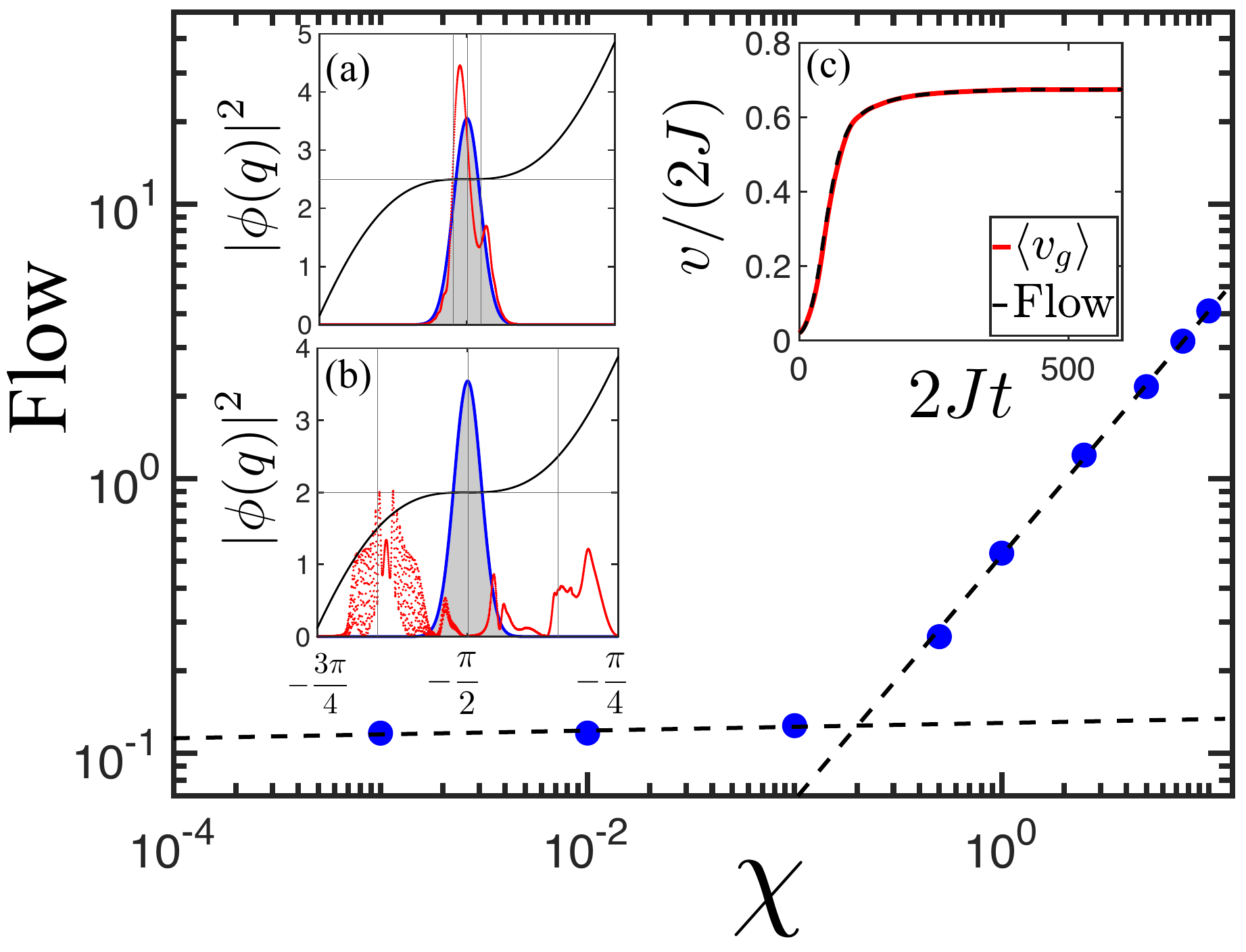}
\caption{\label{Fig:Nonlinearity1}Main figure: a stationary value of flow as a function of nonlinearity. Insets: (a) population numbers at $t = 0$ (blue solid line), and $t\gg0$ (red dots), vs wavenumber $q$, for $\chi = 0.1$; (b)  same as (a) for $\chi = 10$; (c) Flow (black dashed line), average group velocity (red solid), $\langle v_{g}\rangle$ as functions of time for $\chi = 10$. The transition between the SIP and ballistic regimes occurs when the wavepacket in $q$-space ``spills out'' of the initial Gaussian peak, between $\chi = 0.1$ and $\chi = 0.5$.}
\end{figure}

Furthermore, by using the terminology ``weak,'' we imply that the nonlinearity could be treated perturbatively, i.~e. the linear eigenmode representation still provides a valid basis. {This can be guaranteed by ensuring the nonlinear energy contribution to the total internal energy ${\cal H}\equiv H+{\chi\over 2}\sum_n\left|\psi_n\right|^4$ is small compared to the linear energy $H$, given by \cref{Eq:Hamiltonian1D}. We have also compared the average group velocity, which is derived from the linear dispersion relation, with the flow in the presence of the nonlinearity, see \cref{Fig:Nonlinearity1}~(c). Examination of both quantities has shown us that even for the largest values of $\chi$ used in our analysis below, the nonlinear effects (in this $\chi$-range) can be indeed treated as a perturbation to the linear dynamics.} 

Technically, the dynamical equations could be rescaled to fix $\chi\equiv 1$, as only $(\chi/2)\sum_{n}|\psi_{n}|^{4}$ contributes to the total internal energy {${\cal H}$}. In physical photonic networks the nonlinear contribution is governed by the incoming optical power $\mathcal{P} = \sum_n P_n(t = 0)$ rather than changes in the material properties. However,  for theoretical analysis it is convenient to  vary $\chi$ as the relevant parameter in different simulations, while keeping incident power constant.

In general, nonlinear effects that impact the dynamics in periodic systems are expected to emerge. As the nonlinearity provides a mechanism of wave mixing, the initial wavepacket in $q$-space does not remain constant. Rather, the underlying four-wave mixing causes a smearing and splitting of the wavepacket in $q$-space, introducing additional Bloch states to the wavepacket propagation which can change the flow with respect to the underlying linear system. Here we investigate some of the possible nonlinear effects by introducing modifications to the model in \cref{Eq:DynamicsX-spaceLinear}.    

In \cref{Fig:Nonlinearity1} we see how the presence of nonlinearity affects the flow. The distinct crossover towards ballistic propagation between $\chi = 0.1$ and $\chi = 0.5$ is caused by spreading of the wavepacket in $q$-space. Indeed, while at $\chi = 0.1$ the wavepacket remains confined in vicinity of the SIP at $\bar{q} = -\pi/2$,  \cref{Fig:Nonlinearity1}~(a), at higher values of nonlinearity the states corresponding to sufficiently nonzero group velocities become populated (for example, see panel (b) for $\chi = 10$). This explanation is in agreement with \cref{Eq:FlowGroupVel}  and \cref{Fig:Nonlinearity1}~(c). 

In the linear system, the propagation speed in presence of an SIP depends on the wavepacket width: in the hypothetical case of a $q$-space delta-function initial condition, $\sigma\to 0$, the signal won't propagate  as the group velocity is identically zero, however,  broadening the wavepacket introduces proximal states whose group velocities are small but not entirely vanishing. {We restrict our analysis to bandwidths within a range that is  large enough for the  initial wave to be a localized packet in direct space, and small enough to keep the Bloch states in vicinity of the SIP sufficiently populated. For instance, for a  typical system size $N\sim 10^{3}$ used in our simulations, we achieved this balance by choosing $\sigma\sim0.1$, effectively preparing the initial wavepacket $\Delta n\sim 10$ sites wide.The precise shape of the wavepacket is not important as long as this range can be maintained. The Gaussian profile is a reasonable choice for its analytical properties and physical accessibility. The nonlinear effect causes a crossover to the ballistic transport regime as the Bloch-mode population becomes independent of the initial preparation.}     

\section{Protocols for controlling propagation direction \label{Sec:Protocols}} 
\begin{figure}[htbp]
\center\includegraphics[scale = .35]{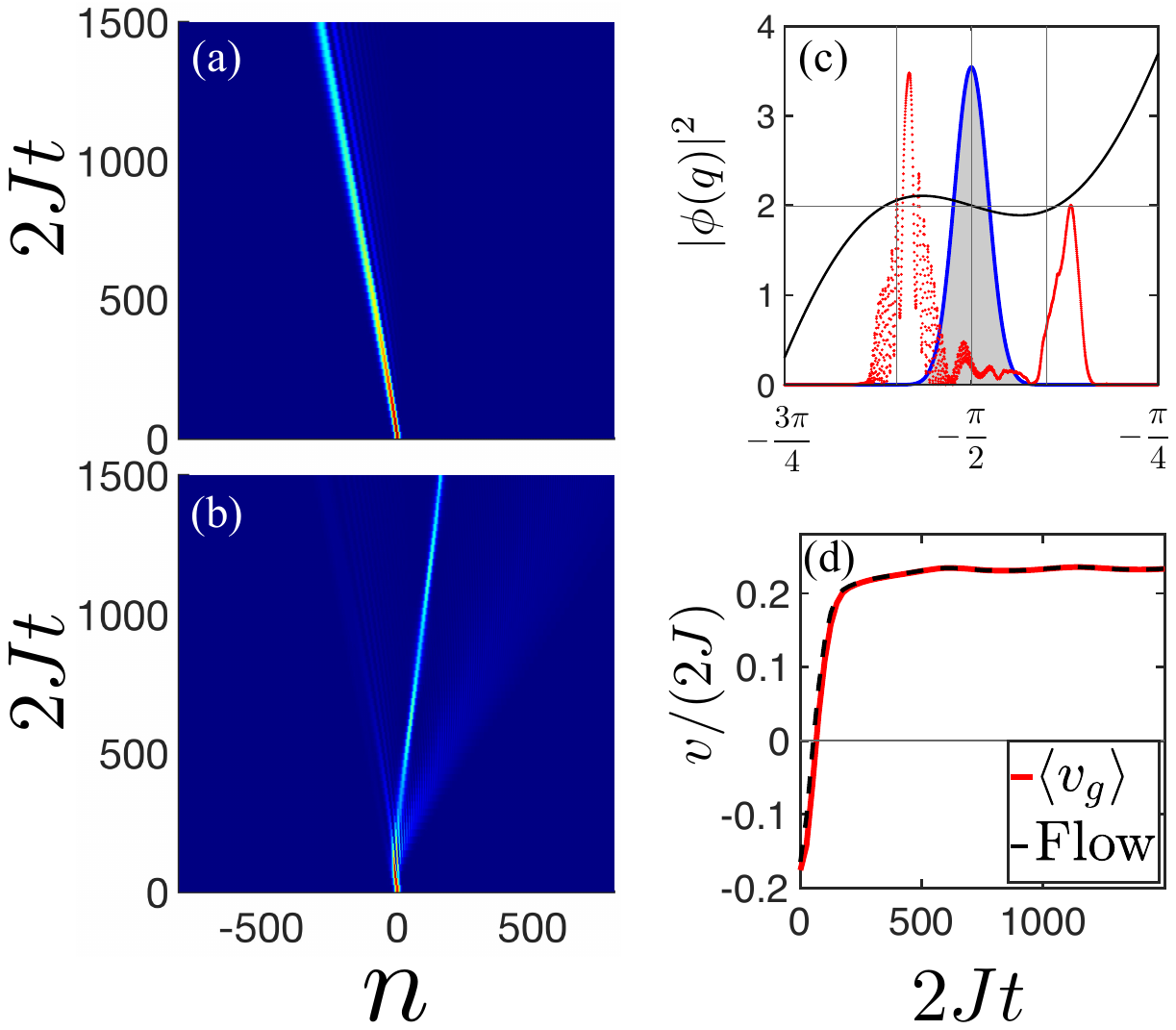}
\caption{\label{Fig:Nonlinearity2}Signal deflection by incident power. (a) Signal profile in a linear system with roton dispersion relation. (b) Signal profile in the same system with nonlinearity $\chi = 5$. (c) Occupation numbers at $t = 0$ (blue solid line), and $t\gg0$ (red dots), vs wavenumber $q$, for $\chi = 5$, black solid line corresponds to the roton dispersion relation; one can see the positive group velocity states become populated. (d) Average group velocity (red solid) and the flow (black dashed) vs time. The initial negative values switch to positive in the stationary regime by the spreading of the wavepacket in $q$-space.}
\end{figure}

\textit{Roton dispersion induced by SIP management.} The dependence of the flow on the population numbers in $q$-space gives an idea how to control not only the signal speed, but also the direction via the manipulation of its incident power. Consider again the dispersion relation \cref{Eq:SIPspectrum}. When $J_{3} > J/3$, the inflection point $q = -\pi/2$ has a negative slope, as it is positioned between a local maximum (to its left) and a local minimum (to its right), creating the so-called roton dispersion relation \cite{Landau1941}.  Hence, in the linear system the initial preparation of a Gaussian packet centered at $q = -\pi/2$ will be followed by the energy propagating in the negative direction. However, as the nonlinearity exceeds some threshold value, the initial Gaussian in $q$-space splits and spreads, exciting states with predominantly  positive group velocity, so that $\langle v_{g}\rangle > 0$, turning the energy flow to the opposite direction. As one can see in \cref{Fig:Nonlinearity2} this effect takes place in a stationary regime, after the time required for the wavepacket to spread in $q$-space.

{A possible experimental realization of the nonlinear pulse propagation we are stusdying at involves arrays of coupled resonators of optical waveguides (CROW). The quantitative details of the coupled array (see \cref{Fig:IntroFigure}~(a) could differ from those in our CMT model, but as long as the Bloch dispersion relation displays an SIP or associated roton behavior, and the Kerr nonlinearity is strong enough, we have every reason to believe that the predicted effects can be produced experimentally. As stated above, the conclusions about the qualitative features of the SIP-related frozen mode regime is expected to remain the same regardless of the specific physical platform. An alternative platform for the realization of such beam dynamics has recently emerged in the frame of acoustic metamaterials where the roton dispersion has been designed via appropriate nonlocal couplings \cite{Chen2021, Wang2022,Bossart2023}.} 

\textit{Nonlinear coupling}. Another possible method of signal deflection is based on a modification of the dispersion relation by nonlinearity. In the one-channel model,  the onsite nonlinearity may cause a vertical shift of the dispersion relation but not deformation of the band. In systems with more complex unit cell structure, uniform onsite nonlinearities can alter relative onsite optical potentials between propagation channels, which deforms the effective dispersion relation. One can still demonstrate this phenomenon in the framework of a one-channel model by introducing nearest-neighbor nonlinear coupling, $J \to J(1 + \mu |\psi_{n}|^{2})$, into \cref{Eq:DynamicsX-spaceLinear}. {Implementation of such non-local nonlinearities has been already reported in electronic circuits, see for example \cite{Marquie1995, Selim2023}. Irrespective, the goal of the present section is to demonstrate the nonlinear dispersion effect with minimal modifications to our relatively simple mathematical model.}

\begin{figure}[htbp]
\center\includegraphics[scale = .3]{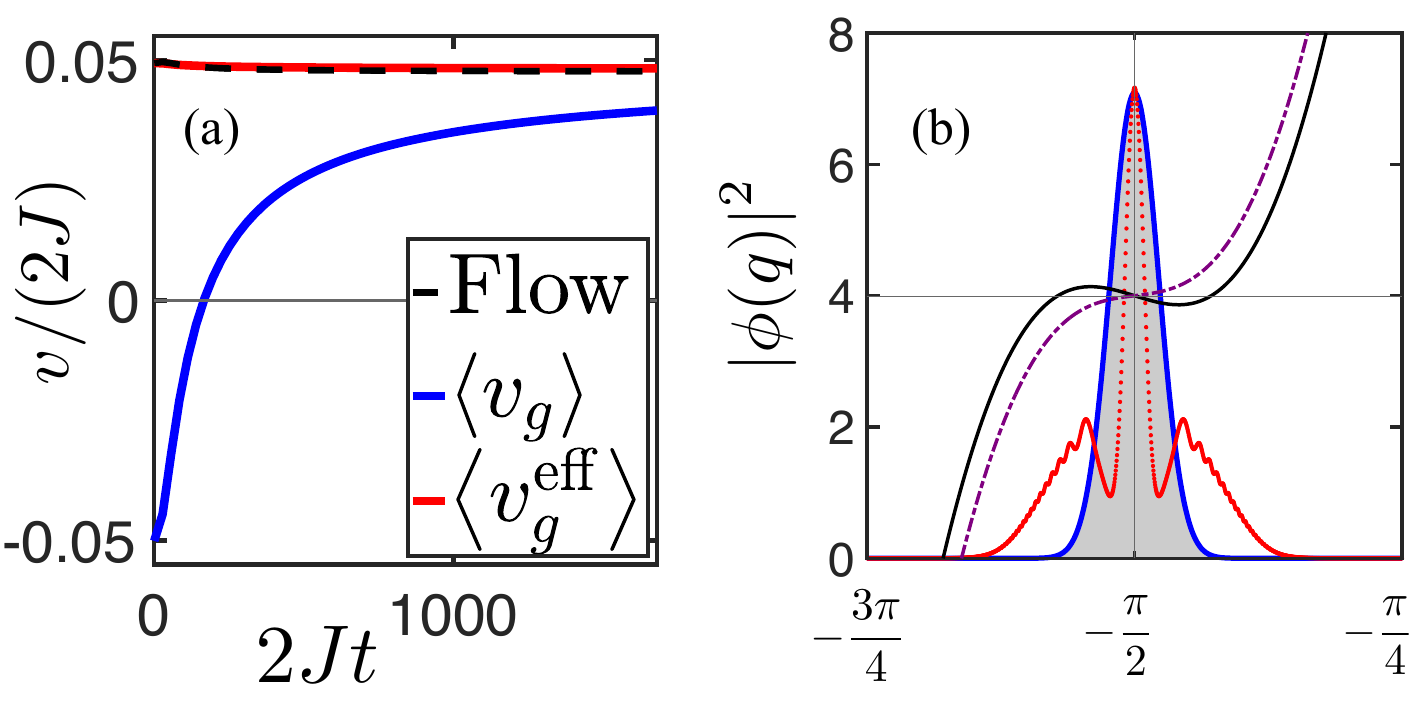}
\caption{\label{Fig:NonlinearCoupling}(a) Flow (black dashed line) for $\mu = 5$ and $J/(3J_{3}) = 0.93$, average group velocity of the underlying roton dispersion relation (blue solid) $\langle v_{g}\rangle$, and the effective group velocity (red solid), $\langle v_{g}^{{\rm eff}}\rangle$, vs time. (b) Occupation numbers at $t\sim 0$ (blue solid line) and at $t\gg 0$ (red dots); black solid line corresponds to the roton dispersion relation, purple dashed-dotted line is the effective curve caused by the coupling nonlinearity at $t\sim 0$.}
\end{figure}

In \cref{Fig:NonlinearCoupling}~(a) one can see that the flow (black dashed line) is positive from the beginning of the dynamical evolution, while the average group velocity (blue solid) is negative as one expects for the underlying roton linear system. Strictly speaking, \cref{Eq:FlowGroupVel} is not applicable as the nonlinearity cannot be treated perturbatively and the dispersion relation is not well-defined. However, \cref{Eq:SIPspectrum} may be conditionally restored for any time step if the coupling parameter $J$ is replaced by $J_{\rm eff}(t) = J[1 + \mu \delta(t)]$, where $\delta(t) \sim \langle|\psi_{n}(t)|^{2}\rangle$. Then, even though $J/(3J_{3}) < 1$ (a condition for a roton dispersion relation), the effective group velocity, $v_{g}^{{\rm eff}}(t) = \partial\omega^{\rm eff}(q, t)/\partial q$, in vicinity of the inflection point at $q = -\pi/2$ will remain positive whenever $\delta(t) > 1 - J/(3J_{3})$. One can see in \cref{Fig:NonlinearCoupling}~(a)  average values of the effective group velocity (red solid), $\langle v_{g}^{{\rm eff}}(t)\rangle$, which is in agreement with the flow.  

Apparently, this effect is only observable  in the short time range, as for $t\gg 0$ the value of $\langle|\psi_{n}|^{2}\rangle \propto 1/N$. Thus $\delta(t)$ becomes negligible and the dispersion relation converges to the roton profile. On the large time scale the direction of signal propagation is governed by the wavepacket distribution in $q$-space. The competition between these two processes may be clarified by \cref{Fig:NonlinearCoupling}~(b): at $t\sim 0$, the propagation is governed by the narrow Gaussian peak (solid blue line) probing the effective $\omega^{{\rm eff}}(q)$ (purple dash-dotted line), at $t\gg 0$ the band is restored toward the roton dispersion relation (black solid curve) while the wavepacket (red dots) probes the states outside the negative group velocity region. 

\section{Conclusions.} In this study, we have developed a theoretical framework for linear and nonlinear dynamics of wavepackets centered at an SIP. In the linear regime, such pulses do not propagate ballistically, due to the zero group velocity at the SIP frequency. We have demonstrated that nonlinearity can result in ballistic propagation of SIP-centered pulses, with the speed and even direction of propagation essentially dependent on the pulse amplitude. {This unique feature that emerges from the interplay between an SIP and nonlinearity provides exciting opportunities for control and manipulation of electromagnetic and acoustic pulses injected into a composite structure that supports an SIP. One possible application is the development of a novel type of beam power router. Another application is to use this unique effect for MW and optical limiting, in which case only pulses with the amplitude below a certain threshold will be transmitted by the structure, while the input pulses with their  amplitude exceeding the threshold will be reflected back. Yet a third application is in a “non-resonant Q-switch,” which prevents radiation leaking from a system unless the pulse amplitude exceeds a threshold value. In all cases, a combination of the enhanced amplitude (see, e.g., Ref. \cite{Figotin2011} and references therein) of the frozen mode and the enhanced response to nonlinearities in the vicinity of an SIP provide great flexibility in achieving the desirable threshold values}.

\textit{Acknowledgments.} We acknowledge partial support from DEC, DE-SC0024223,
NSF-EFMA 161109, the Simons Foundation MPS-733698,
BSF2022158, and AFOSR LRIR 21RYCOR019.


\appendix
\section{Integral Evaluation\label{Sec:IntegralAppendix}}
In this section we provide a detailed, though not rigorous, evaluation of integral which appears in \cref{Eq:psinstep2}, i.~e.
\begin{equation}\label{Eq:Izeps}
\mathcal{I}(z, \epsilon) = \frac{1}{\pi}\int\limits_{0}^{\infty}dx e^{-\epsilon x^{2}}\cos\left(\frac{x^{3}}{3} - zx\right).
\end{equation}
First, we notice that $\mathcal{I}(z, 0) = \mathrm{Ai}(-z)$, the Airy function, which is a solution of the differential equation \cite{AbramowitzStegun}
\begin{equation}\label{Eq:AiryEquation}
y^{\prime\prime} + zy = 0.
\end{equation} 
Noticing that 
\[
\frac{\partial^{2k}}{\partial z^{2k}}\cos\left(\frac{x^{3}}{3} - zx\right) = (-1)^{k}x^{2k}\cos\left(\frac{x^{3}}{3} - zx\right),
\]
and expanding $e^{-\epsilon x^{2}}$ into the Taylor series for any $\epsilon>0$, we get
\begin{widetext}
\begin{multline}\label{Eq:IntegralSteps}
\mathcal{I}(z, \epsilon) = \frac{1}{\pi}\int\limits_{0}^{\infty}dx e^{-\epsilon x^{2}}\cos\left(\frac{x^{3}}{3} - zx\right) = \frac{1}{\pi}\int\limits_{0}^{+\infty}dx \sum\limits_{k = 0}^{\infty}\frac{(-1)^{k}\eps^{k}}{k!}x^{2k}\cos\left(\frac{x^{3}}{3} - zx\right) =\\
 \frac{1}{\pi}\int\limits_{0}^{+\infty}dx \sum\limits_{k = 0}^{\infty}\frac{\eps^{k}}{k!}\frac{\partial^{2k}}{\partial z^{2k}}\cos\left(\frac{x^{3}}{3} - zx\right) = \sum\limits_{k = 0}^{\infty}\frac{\eps^{k}}{k!}\frac{\partial^{2k}}{\partial z^{2k}}\mathcal{I}(z, 0) = \sum\limits_{k = 0}^{\infty}\frac{\eps^{k}}{k!}\frac{\partial^{2k}}{\partial z^{2k}}\mathrm{Ai}(-z).
\end{multline}
\end{widetext}

Introduce notations $F(z) = \mathrm{Ai}(-z)$, $G(z) = \frac{d}{dz}\mathrm{Ai}(-z)$,
then
\begin{equation}\label{Eq:AiryDer}
\begin{split}
&F^{(2)}(z) = -zF(z)\\
&F^{(4)}(z) = z^{2}\left[-2z^{-2}G(z) + F(z)\right]\\
&F^{(6)}(z) = z^{3}\left[4z^{-3} + 6z^{-2}G(z) - F(z)\right]\\
&F^{(2k)}(z) = z^{k}\left[\hdots + (-1)^{k}F(z)\right].
\end{split}
\end{equation} 
 First, consider positive values of $z$. For $z > 1$ the strongest order of $z$ in the asymptotic approximation of the Airy function, $\mathrm{Ai}(-z)$,  is  
 \begin{equation}\label{Eq:AiryApprox1}
 \begin{split}
 &\mathrm{Ai}(-z) \propto z^{-1/4}\sin\left(\frac{2}{3}z^{3/2} + \frac{\pi}{4}\right),\\
 &\mathrm{Ai}^{\prime}(-z) \propto z^{1/4}\cos\left(\frac{2}{3}z^{3/2} + \frac{\pi}{4}\right).
\end{split}
 \end{equation}
Therefore, the $k$-th expression of \cref{Eq:AiryDer} is actually
\[
F^{(2k)}(z) = z^{k-1/4}\left[\mathcal{O}(z^{-3/2}) + (-1)^{k}z^{1/4}F(z)\right],
\]
$k \geqslant 2$, $z^{1/4}F(z)\sim 1$.
For $z < 0$, Airy function is not periodic, but quickly decaying:
 \begin{equation}\label{Eq:AiryApprox2}
 \begin{split}
&\mathrm{Ai}(-z) \propto |z|^{-1/4}\exp\left\{-\frac{2}{3}|z|^{3/2}\right\},\\ 
&\mathrm{Ai}^{\prime}(-z) \propto |z|^{1/4}\exp\left\{-\frac{2}{3}|z|^{3/2}\right\},
\end{split}
\end{equation}
 and 
 \begin{multline*}
F^{(2k)}(z) = |z|^{k-1/4}e^{-\frac{2}{3}|z|^{3/2}}\Bigl[\mathcal{O}(|z|^{-3/2}) +\\ |z|^{1/4}e^{+\frac{2}{3}|z|^{3/2}}F(z)\Bigr], \quad |z|^{1/4}e^{+\frac{2}{3}|z|^{3/2}}F(z)\sim 1.
\end{multline*}
Therefore we may approximate the derivatives as:
\[
\frac{\partial^{2k}}{\partial z^{2k}}\mathrm{Ai}(-z) \approx (-1)^{k}z^{k}\mathrm{Ai}(-z),
\]
and plugging it into \cref{Eq:IntegralSteps} we obtain
\begin{equation}\label{Eq:Solution}
\mathcal{I}(z, \epsilon)\approx \sum\limits_{k = 0}^{\infty}\frac{\eps^{k}}{k!}(-1)^{k}z^{k}\mathrm{Ai}(-z) = e^{-\epsilon z}\mathrm{Ai}(-z).
\end{equation}

We have to make one remark about this derivation: though integral \cref{Eq:Izeps} converges for any $\epsilon\geqslant 0$, the last step in \cref{Eq:IntegralSteps}, a change of integration and summation in their order, is not rigorous justified, as convergence is not guaranteed for any value of $\epsilon$. Actually, while the integral \cref{Eq:Izeps} converges  quicker for larger $\epsilon$, the sum converges better for $\epsilon < 1$. There is no contradiction here, it is a choice of the approximation domain. 
The parameters $\epsilon$, $z$ are not independent as they are introduced via physical variables
\[
\epsilon = (1/2)(\alpha t)^{-2/3}\sigma^{-2}, \quad z = n(\alpha t)^{-1/3},
\]
 in \cref{Eq:psinstep2} of the main text. To satisfy the initial condition, the integral {\cref{Eq:Izeps} should behave as $\mathcal{O}\left[t^{1/3}\right]$ for $t\to 0$. It is apparently not the case for \cref{Eq:Solution}. It only means that this approximation is not valid for $t\to 0$. Technically speaking, the time domain of guarantied applicability is
\[
\sigma^{-3}\ll \alpha t \ll n^{3},
\] 
quite a realistic range. Practically, one can see in comparison with the numerical simulations that the approximation qualitatively captures all the phenomena associated with SIP dynamics in almost the entire time domain.

\section{Flow in presence of SIP\label{Sec:FlowAppendix}}
In absence of losses we define flow as 
\begin{equation}
\mathcal{F}(t) \overset{\rm def}{=} \sum_{n}n\frac{d}{dt}\left|\psi(t, n)\right|^{2} = \frac{d}{dt}\sum_{n}n|\psi(t, n)|^{2}.
\end{equation}
Using the explicit expression for signal propagation in presence of the SIP at $q = -\pi/2$ (Eq.~(6) of the main text) one can write 
\begin{widetext}
\begin{multline}
\sum_{n}n|\psi(t, n)|^{2} 
=
  2\left(\frac{\pi}{\sigma^{2}}\right)^{1/2}\left(\alpha t\right)^{-2/3}\int\limits_{-\infty}^{+\infty} dx x e^{-\frac{x}{\sigma^{2}\alpha t}}\mathrm{Ai}^{2}\left[- x\left(\alpha t\right)^{-1/3}\right] 
  =
2\left(\frac{\pi}{\sigma^{2}}\right)^{1/2}\int\limits_{-\infty}^{+\infty} dy y e^{-\frac{y}{\sigma^{2}(\alpha t)^{2/3}}}\mathrm{Ai}^{2}\left(- y\right) 
=\\ 
-2\left(\frac{\pi}{\sigma^{2}}\right)^{1/2}\frac{\partial}{\partial a}\left[\int\limits_{-\infty}^{+\infty} dy e^{-a y}\mathrm{Ai}^{2}\left(- y\right)\right]
=
-2\left(\frac{\pi}{\sigma^{2}}\right)^{1/2}\frac{\partial}{\partial a}\left[\frac{e^{a^{3}/12}}{2\sqrt{\pi a}}\right]
 \xrightarrow{t\gg \alpha(2\sigma)^{-3/2}} \frac{\sigma^{2}\alpha t}{2},
\end{multline}
\end{widetext}
where $a = \sigma^{-2}(\alpha t)^{-2/3}$.
Therefore,
\begin{equation}
\mathcal{F}(t) \xrightarrow{t\gg \alpha(2\sigma)^{-3/2}} \frac{\sigma^{2}\alpha}{2}.
\end{equation}
This result can also be obtained using the equality of the flow to the average group velocity, $\mathcal{F}(t) = \langle v_{g}\rangle$:
\begin{widetext}
\begin{multline}
\langle v_{g}\rangle 
= 
\int\limits_{-\pi}^{+\pi} dq \left(\frac{\partial\omega}{\partial q}\right)|\phi(q)|^{2}
\approx
  -\frac{2J}{\sqrt{\pi\sigma^{2}}}\int\limits_{-\infty}^{+\infty}dq \left[\sin q + \sin 3q\right]\exp\left\{-\frac{(q-\bar{q})^{2}}{\sigma^{2}}\right\}\approx\\ 
 \frac{2J}{\sqrt{\pi\sigma^{2}}}\int\limits_{-\infty}^{+\infty}dp \left[-\frac{p^{2}}{2} + \frac{9p^{2}}{2}\right]\exp\left\{-\frac{p^{2}}{\sigma^{2}}\right\} 
 = 
\frac{\alpha}{\sqrt{2\pi(\sigma/\sqrt{2})^{2}}}\int\limits_{-\infty}^{+\infty}dp p^{2}\exp\left\{-\frac{p^{2}}{2(\sigma/\sqrt{2})^{2}}\right\} 
 =
 \frac{\alpha\sigma^{2}}{2},
\end{multline}
\end{widetext}
where $\sin q$, $\sin3q$ are approximated in vicinity of SIP, $\bar{q} = -\pi/2$.

\section{Flow and average velocity equality\label{Sec:FlowVgProofAppendix}}
The proof of \cref{Eq:FlowGroupVel} of the main text, $\mathcal{F}(t) = \langle v_{g}(q, t)\rangle$, is straightforward:
\begin{widetext}
\begin{multline}
\mathcal{F}(t) 
=
 \frac{d}{dt}\int dx x|\psi(x, t)|^{2} 
 =
  \frac{d}{dt} \iint dq dp \int dx x \phi^{\ast}(q, t)\phi(p, t)e^{-i(q - p)x} 
=\\
 \frac{d}{dt} \iint dq dp \int dx x e^{i(p - q)x}e^{-i[\omega(p) - \omega(q)]t}C^{\ast}(q, t)C(p, t),
\end{multline}
\end{widetext}
where $C(q, t)$ is a slow function of time, in the linear system a constant. One can proceed further as
\begin{widetext}
\begin{multline}
\mathcal{F}(t) = \frac{d}{dt} \iint dq dp e^{-i[\omega(p) - \omega(q)]t}C^{\ast}(q, t)C(p, t)(-i)\frac{\partial }{\partial p}\int dx e^{i(p - q)x} 
=\\
 \frac{d}{dt} \iint dq dp e^{-i[\omega(p) - \omega(q)]t}C^{\ast}(q, t)C(p, t)(-i)\frac{\partial }{\partial p}\delta(p - q) 
 =
 i\frac{d}{dt} \int dq e^{i\omega(q)t}C^{\ast}(q, t)\frac{\partial }{\partial q}\left[e^{-i\omega(q)t}C(q, t)\right], 
\end{multline}
\end{widetext}
where we use the equality 
\[
\int dx f(x)\frac{\partial }{\partial x}\delta(x - x_{0}) = -f^{\prime}(x_{0}). 
\]
Hence,
\begin{widetext}
\begin{multline}
\mathcal{F}(t) = \frac{d}{dt} \int dq\left[\left(\frac{\partial\omega}{\partial q}\right)|C(q, t)|^{2} + iC^{\ast}(q, t)\frac{\partial C}{\partial q}\right] 
= \\
\int dq\left(\frac{\partial\omega}{\partial q}\right)|\phi(q, t)|^{2} + \int dq\left(\frac{\partial\omega}{\partial q}\right)t\frac{d}{dt}|\phi(q, t)|^{2} + \frac{i}{2}\frac{d}{dt}\int dq |\phi(q, t)|^{2}. 
\end{multline}
\end{widetext}
The last term is always equal to zero due to norm conservation,
\[
\frac{d}{dt}\int dq |\phi(q, t)|^{2} = 0.
\]
The second term is exactly equal  to zero in the linear system as $\dot{C_{0}}\equiv 0$. At $\chi\neq 0$ the second term  is still negligible. Indeed, the norm exchange between the Bloch modes slows down by approaching stationary regime, so  $t\frac{d}{dt}|\phi(q, t\to \infty )|^{2}\to 0$. The norm exchange rate, $\frac{d}{dt}|\phi(q)|^{2}$, is nonzero at $t\to 0$ only, which makes $t\frac{d}{dt}|\phi(q, t)|^{2}\Big|_{t\to 0}\to 0$ as well. Finally,
\[
\mathcal{F}(t) = \int dq\left(\frac{\partial\omega}{\partial q}\right)|\phi(q, t)|^{2} = \langle v_{g}\rangle.
\]   


%
\end{document}